\documentclass[aps,prb,preprint]{revtex4}

\usepackage{graphicx}
\begin{document}

\title{Quantum transport through a graphene nanoribbon-superconductor junction}

\author{Qing-feng Sun$^{1,\star}$ and X. C. Xie$^{2,1}$ }
\affiliation{$^1$Beijing National Lab for Condensed Matter Physics
and Institute of Physics, Chinese Academy of Sciences, Beijing
100190, China\\
$^2$Department of Physics, Oklahoma State University, Stillwater,
Oklahoma 74078 }

\date{\today}

\begin{abstract}
We study the electron transport through a graphene
nanoribbon-superconductor junction. Both zigzag and armchair edge
graphene nanoribbons are considered, and the effects of the magnetic
field and disorder on the transport property are investigated. By
using the tight-binding model and the non-equilibrium Green's
function method, the expressions of the current, conductance, normal
tunneling coefficient, and Andreev reflection coefficient are
obtained. For a clean system and at zero magnetic field, the linear
conductance increases approximatively in a linear fashion with the
on-site energy. In the presence of a magnetic field and a moderate
disorder, the linear conductance exhibits plateau structures for
both armchair and zigzag edges. The plateau values increase with the
width of the graphene ribbon. With a wide sample width, a saturated
plateau value of $|\nu|e^2/h$ emerges at the filling factor $\nu$.
For a small filling factor, the conductance can reach the saturated
value at a small width, but for a high filling factor, it requires
to have a quite wide sample width to reach the saturated value. In
particular, the Andreev reflection coefficient is always at $0.5$
after reaching the saturated value, independent of any system
parameters. In addition, we also consider the finite bias case, in
which the Andreev reflection coefficient and normal tunneling
coefficient are studied.
\end{abstract}


\maketitle

\section{Introduction}

The recent experimental realization of graphene,\cite{ref1,ref2} a
single-layer carbon atoms arranged in a honeycomb lattice, has
generated a great attention in the condensed matter
community.\cite{ref3,ref4} Graphene has an unique band structure
with a linear dispersion relation of the low-lying excitations,
which leads to many peculiar properties,\cite{ref3,ref4} such as its
quasi-particles obeying the Dirac-like equation and having the
relativistic-like behaviors with zero rest mass, the Hall plateaus
having the half-integer values $g (n+1/2) e^2/h$ with the degeneracy
$g=4$. For the neutral graphene, its Fermi level passes through the
Dirac points, the six corners of the hexagonal first Brillouin zone.
By varying the gate voltage, the charge carriers of graphene can be
easily tuned experimentally. Then the Fermi level can be above or
below the Dirac points.

Very recently, some works have begun to investigate the transport
behaviors of the graphene-superconductor
junctions.\cite{ref5,ref6,ref7,ref8,ref9,ref9n,ref10,ref11,ref12,ref13,ref14,aref1,aref2}
While a metal coupled to a superconductor, the Andreev reflection
occurs in the interface between the metal and
superconductor,\cite{ref15} in which the interface reflects an
electron incident from the normal metal side as a hole and a Cooper
pair is created in the superconductor. For a bias below the
superconductor gap, the Andreev reflection determines the
conductance of the metal-superconductor junction since the normal
tunneling cannot occur. In the usual metal-superconductor junction,
the Andreev reflected hole retraces the path of the incident
electron, so this Andreev reflection is also called Andreev
retroreflection. But for the graphene-superconductor junction,
Beenakker recently found that a new kind of reflection (specular
Andreev reflection) occurs while the incident electron and reflected
hole are at the conduction and valence bands,
respectively.\cite{ref5} Afterwards, many papers have studied the
graphene and superconductor hybrid system, including the
graphene-based normal-superconductor
(N-S),\cite{ref5,ref6,ref7,ref13} S-N-S,\cite{ref8,ref9,ref9n}
S-insulator-S,\cite{ref12} S-ferromagnet-S,\cite{ref10} and etc.
Several other effects due to the coupling of graphene and
superconductor, such as Josephson effect\cite{ref8,ref9,ref12} and
multiple Andreev reflection processes,\cite{ref14} have been
theoretically analyzed. On the experimental side, good contacts
between the superconductor electrodes and graphene have been
realized,\cite{ref16,ref17} and the Josephson current through a
S-graphene-S junctions has been measured. A finite supercurrent was
observed at zero charge density.

In this paper, we carry out theoretical study of the transport
characteristics of a graphene nanoribbon-superconductor junction.
There are four new aspects beyond the previous studies: (i) We
study the system consisting of a graphene nanoribbon with a finite
width coupled to the superconductor electrode. The previous
theoretical papers only consider the infinite-wide
graphene-superconductor junction or a graphene strip between two
superconductor leads with the strip width much larger than the
strip length. On the experiment side, the graphene nanoribbon has
been successfully fabricated, and the width of the nanoribbon can
be in the order of ten or sub-ten nanometers.\cite{ref18} (ii) In
our model, the graphene nanoribbon is directly coupled to the
superconductor electrode, and the incident electrons from the
graphene are allowed to enter into the superconductor
electrode as the Cooper pairs. In the previous papers, those
authors only considered a pair potential in the graphene induced
by depositing of a superconductor electrode on top of the graphene
sheet. (iii) We consider a perpendicular magnetic field applied to
the graphene, as was done in a recent experiment.\cite{ref16} On
the superconductor side, the magnetic field vanishes due to the
Meissner effect. (iv) The effect of disorder on the transport
property is investigated since in a real graphene sample, the
disorder is always there to a certain degree. In fact, in the
previous studies, the effects of disorder and magnetic field are
thus far neglected.

By using the tight binding
model and the nonequilibrium Green function method, the current and
Andreev reflection coefficient are obtained. Both zigzag edge and
armchair edge graphene nanoribbons are considered. For the zigzag
edge and at a zero magnetic field, the linear conductance exhibits
step structures for the narrow graphene ribbon. With a magnetic
field, the conductance depends strongly on the system parameters. In
the presence of disorder, the linear conductance shows plateaus at a
high magnetic field. On the other hand, for the armchair edge, a
zero-conductance region emerges because of the existence of an
energy gap in the graphene nanoribbon. This zero conductance is
robust against disorder. In addition, we also consider a finite-bias
case, in which the Andreev reflection coefficient and normal
tunneling coefficient are investigated.

The rest of the paper is organized as follows. In Section II, the
model for graphene nanoribbon-superconductor junction is presented
and the formalisms for calculating the current and the Andreev
reflection coefficient are derived. In Section III and IV, we study
the linear conductance and the transport with a finite bias,
respectively. Finally, a brief summary is presented in Section V.

\section{model and formalism}

We consider the system consisting of a graphene nanoribbon coupled
to a superconductor lead (as shown in Fig.1) with the Hamiltonian:
\begin{eqnarray}
 H = H_{G} + H_{S} + H_{C},
\end{eqnarray}
where $H_{G}$, $H_{S}$, and $H_{C}$ are the Hamiltonians of the
graphene region, superconductor lead, and coupling of the graphene
and superconductor lead, respectively. For a semi-infinite graphene
nanoribbon, $H_G$ in the tight-binding representation is of the
form:\cite{ref19,ref20}
\begin{eqnarray}
 H_{G} =\sum_{i,\sigma} \epsilon_i a^{\dagger}_{i\sigma} a_{i\sigma}
      -\sum_{<ij>,\sigma} t e^{i\phi_{ij}} a_{i\sigma}^{\dagger} a_{j\sigma},
\end{eqnarray}
where $a_{i\sigma}$ and $a_{i\sigma}^{\dagger}$ are the
annihilation and creation operators at the discrete site $i$, and
$\epsilon_i$ is the on-site energy which can be controlled by the
gate voltage in an experiment. Two kinds of edges, zigzag and
armchair, are considered (see Fig.1a and 1b). The graphene ribbon
is divided into two regions. The left-side of the semi-infinite
region is without disorder and $\epsilon_i=E_L$ there. The
disorder exists only in the center region of the
graphene-nanoribbon (see the box with the dotted-line in Fig.1).
Here we consider the on-site disorder causing by the nonmagnetic
impurities or by the random potential difference of the substrate.
Due to the disorder, the on-site energy $\epsilon_i = E_L +w_i$,
where $w_i$ is the on-site disorder energy and $w_i$ is uniformly
distributed in the range $[-W/2, W/2]$ with $W$ being the disorder
strength. The size of the disorder region is described by the
width $N$ and length $L$. In Fig.1a and 1b, $N=3$, $L=4$ and
$N=4$, $L=2$, respectively. The second term in Eq.(2) is the
nearest neighbor hopping. When the graphene ribbon is under a
uniform perpendicular magnetic field $B$, a phase $\phi_{ij}$ is
added in the hopping elements, and $\phi_{ij}=\int_i^j \vec{A}
\cdot d\vec{l}/\phi_0$ with the vector potential
$\vec{A}=(-By,0,0)$ and $\phi_0=\hbar/e$.

Experimentally, it is possible to have the superconductor electrode
in a good contact with the graphene.\cite{ref16} The electrons from
the graphene can easily enter into the superconductor electrode as
the Cooper pairs or vice versa. So we consider that the graphene
nanoribbon is directly coupled to the superconductor electrode. The
superconductor electrode is described by a continuum model and it
does not have the honeycomb structure of the graphene. Then the
Hamiltonian $H_{S}$ is:
\begin{eqnarray}
  H_{S} = \sum\limits_{{\bf k},\sigma} \epsilon_{{\bf k}}
   b^{\dagger}_{{\bf k}\sigma} b_{{\bf k}\sigma}
   + \sum\limits_{{\bf k}} \left( \Delta b^{\dagger}_{{\bf
   k}\uparrow} b^{\dagger}_{-{\bf k}\downarrow} + \Delta b_{-{\bf
   k}\downarrow} b_{{\bf k}\uparrow} \right),
\end{eqnarray}
where $b_{{\bf k}\sigma}$ and $b^{\dagger}_{{\bf k}\sigma}$ are the
annihilation and creation operators in the superconductor lead with
the momentum ${\bf k}= (k_x, k_y)$. Here we consider the s-wave
superconductor and $\Delta$ is the superconductor gap. The
superconductor region is without the magnetic field due to the
Meissner effect or that the magnetic field is only added in the
graphene region. The Hamiltonian $H_{C}$ of the coupling between the
superconductor lead and the graphene nanoribbon is:
\begin{eqnarray}
  H_{C} = \sum\limits_{i,\sigma} t_c a^{\dagger}_{i\sigma}
  b_{\sigma}(y_i) +H.c.
\end{eqnarray}
Here only the surface carbon atoms couple to the superconductor
lead, and $y_i$ is the vertical position of the carbon atom $i$.
$b_{\sigma}(y)$ is the annihilation operators at the position
$(0,y)$ of real space, and
\begin{eqnarray}
 b_{\sigma}(y) =\sum\limits_{k_x,k_y} e^{ik_y y} b_{{\bf k}\sigma}
\end{eqnarray}

The current flowing through the graphene nanoribbon-superconductor
junction can be calculated from the evolution of the total number
operator for electrons in the left graphene-nanoribbon
lead,\cite{ref21}
\begin{eqnarray}
 I & =&  -e\langle \frac{d}{dt} \sum\limits_{i\in L,\sigma}
 a^{\dagger}_{i\sigma} a_{i\sigma} \rangle \nonumber\\
 & =& \frac{e}{\hbar} \sum\limits_{i\in L, j\in C} \int
 \frac{d\omega}{2\pi} \left\{ t_{ij} G^<_{ji,11} -t_{ij} G^<_{ij,22}
 -t_{ji} G^<_{ij,11} + t_{ji} G^<_{ji,22} \right\},
\end{eqnarray}
where $t_{ij}=t e^{i\phi_{ij}}$. Here $i\in L$ and $j\in C$
represent that the site index $i$ and $j$ are in the left graphene
lead and center region. ${\bf G}^<_{ij}(\omega)$ is the matrix Green
function in Nambu representation, and it is the Fourier
transformation of ${\bf G}^<_{ij}(t)$:
\begin{eqnarray}
 {\bf G}^<_{ij}(t) = i \left( \begin{array}{ll}
 \langle a^{\dagger}_{j\uparrow}(0) a_{i\uparrow}(t)\rangle &
 \langle a_{j\downarrow}(0) a_{i\uparrow}(t)\rangle \\
 \langle a^{\dagger}_{j\uparrow}(0)
 a^{\dagger}_{i\downarrow}(t)\rangle &
 \langle a_{j\downarrow}(0)
 a^{\dagger}_{i\downarrow}(t)\rangle\end{array}
 \right),
\end{eqnarray}
By using the Dyson equation, the current expression in Eq.(6) can be
rewritten as:
\begin{eqnarray}
  I =\frac{e}{h} \int d\omega Tr \left( \begin{array}{ll} 1 &0\\0 &-1
  \end{array} \right) \bigotimes {\bf I}_{Nc}
  \left\{({\bf \Sigma}^a_L - {\bf \Sigma}^r_L) {\bf G}^< +
   {\bf \Sigma}^<_L ({\bf G}^r - {\bf G}^a) \right\}.
\end{eqnarray}
Here ${\bf G}^{r,a,<}(\omega)$ are the $2Nc \times 2Nc$ matrix
Green's functions in the center region with $Nc$ being the number of
sites in the center region. The retarded and advanced Green's
functions ${\bf G}^{r,a}$ are defined in the standard
way.\cite{ref22} ${\bf I}_{Nc}$ is the $Nc \times Nc$ unit matrix.
${\bf \Sigma}^{r,a,<}_L(\omega)$ are the retarded, advanced, and
lesser self-energies of coupling to the left graphene lead, and they
are:
\begin{eqnarray}
 {\bf \Sigma}^r_{L,ij} & = & \sum\limits_{n\in L,m\in L} \left(
 \begin{array}{ll} t_{in}{\bf g}^r_{nm,11} t_{mj}  &0\\0 & t_{in}^* {\bf g}^r_{nm,22}t_{mj}^*
 \end{array}\right) \\
 {\bf \Sigma}^a_{L} & = &  {\bf \Sigma}^{r\dagger}_{L} \\
 {\bf \Sigma}^<_{L} & = & \left(
 \begin{array}{ll} if_{\uparrow}(\omega){\bf \Gamma}_{L\uparrow}(\omega)  &0\\0
 & if_{\downarrow}(\omega){\bf \Gamma}_{L\downarrow}(\omega)
 \end{array}\right),
\end{eqnarray}
where $f_{\uparrow}(\omega)=f(\omega-eV)$ and
$f_{\downarrow}(\omega)=f(\omega+eV)$ with $V$ being the bias
voltage and $f(\omega)$ being the Fermi distribution function, and
${\bf \Gamma}_{L\uparrow}(\omega) \equiv i({\bf \Sigma}^r_L -{\bf
\Sigma}^a_L)_{11}$ and ${\bf \Gamma}_{L\downarrow}(\omega) \equiv
i({\bf \Sigma}^r_L -{\bf \Sigma}^a_L)_{22}$. ${\bf
g}^r_{nm}(\omega)$ in Eq.(9) is the surface Green's function of the
semi-infinite graphene nanoribbon, that can be numerically
calculated.\cite{ref23} With the aid of the self-energy functions in
Eqs.(9-11), the current $I$ is finally reduced to:
\begin{eqnarray}
 I & =& I_{\uparrow} +I_{\downarrow}, \\
 I_{\uparrow} &=& \frac{ie}{h} \int d\omega Tr{\bf
 \Gamma}_{L\uparrow} \{{\bf G}^< +f_{\uparrow} ({\bf G}^r-{\bf G}^a
 )\}_{11}, \\
  I_{\downarrow} &=& -\frac{ie}{h} \int d\omega Tr{\bf
 \Gamma}_{L\downarrow} \{{\bf G}^< +f_{\downarrow} ({\bf G}^r-{\bf G}^a
 )\}_{22}.
\end{eqnarray}

As shown in the Appendix, the self-energies ${\bf \Sigma}^{r,a,<}_R$
of coupling to the superconductor lead can be obtained by ${\bf
\Sigma}_R^r = -(i/2){\bf \Gamma}_R$, ${\bf \Sigma}_R^a = (i/2){\bf
\Gamma}_R$, and ${\bf \Sigma}_R^< = i f(\omega){\bf \Gamma}_R$. Then
by using the Keldysh equation ${\bf G}^< ={\bf G}^r {\bf \Sigma}^<
{\bf G}^a$, ${\bf G}^r-{\bf G}^a= {\bf G}^r({\bf \Sigma}^r-{\bf
\Sigma}^a) {\bf G}^a$, and the self-energies ${\bf \Sigma}^{r,a,<}
={\bf \Sigma}^{r,a,<}_L +{\bf \Sigma}^{r,a,<}_R$, the currents
$I_{\uparrow}$ and $I_{\downarrow}$  can be rewritten as:
\begin{eqnarray}
 I_{\uparrow} &=& \frac{e}{h} \int d\omega Tr\{ {\bf
 \Gamma}_{L\uparrow} \left[{\bf G}^r {\bf \Gamma}_R {\bf G}^a\right]_{11}
 (f_{\uparrow}-f) + {\bf \Gamma}_{L\uparrow} {\bf G}^r_{12} {\bf \Gamma}_{L\downarrow} {\bf G}^a_{21}
 (f_{\uparrow}-f_{\downarrow}) \}, \\
 I_{\downarrow} &=& -\frac{e}{h} \int d\omega Tr\{ {\bf
 \Gamma}_{L\downarrow} \left[{\bf G}^r {\bf \Gamma}_R {\bf G}^a\right]_{22}
 (f_{\downarrow}-f) + {\bf \Gamma}_{L\downarrow} {\bf G}^r_{21} {\bf \Gamma}_{L\uparrow} {\bf G}^a_{12}
 (f_{\downarrow}-f_{\uparrow}) \}.
\end{eqnarray}
Here $ Tr \{ {\bf  \Gamma}_{L\uparrow} \left[{\bf G}^r {\bf
\Gamma}_R {\bf G}^a\right]_{11} \} \equiv T_{\uparrow}(\omega)$ and
$Tr\{ {\bf  \Gamma}_{L\downarrow} \left[{\bf G}^r {\bf \Gamma}_R
{\bf G}^a\right]_{22} \} \equiv T_{\downarrow}(\omega)$ are the
normal tunneling coefficients for the incident spin-up electron and
spin-down hole with the energy $\omega$, and $Tr\{{\bf
\Gamma}_{L\uparrow} {\bf G}^r_{12} {\bf \Gamma}_{L\downarrow} {\bf
G}^a_{21}\} \equiv T_{A\uparrow}(\omega)$ and $Tr\{ {\bf
\Gamma}_{L\downarrow} {\bf G}^r_{21} {\bf \Gamma}_{L\uparrow} {\bf
G}^a_{12} \} \equiv T_{A\downarrow}(\omega)$ are the Andreev
reflection coefficients. Since the Pauli
matrices $\hat{\sigma}_{x,y,z}$ commute with the Hamiltonian $H$,
the normal transmission coefficients $T_{\uparrow}(\omega) =
T_{\downarrow}(-\omega) \equiv T(\omega)$ and the Andreev reflection
coefficients $ T_{A\uparrow}(\omega) = T_{A\downarrow}(-\omega)
\equiv T_A(\omega)$.

In the following, we need to calculate the Green's functions ${\bf
G}^r$ and ${\bf G}^a$ of the center region. Since the self-energy
${\bf \Sigma}^r$ has been obtained before and by using the Dyson's
equation, the Green's function ${\bf G}^r$ is simply of the form
\begin{equation}
 {\bf G}^r(\omega) = 1/\left( \omega {\bf I}_{2Nc} -{\bf H}_{center}
 -{\bf \Sigma}^r\right),
\end{equation}
and in addition ${\bf G}^a ={\bf G}^{r\dagger}$, where ${\bf
H}_{center}$ is the Hamiltonian of the center region in the Nambu
representation.

In the numerical calculations, we take the hopping energy
$t=t_c=2.75eV$ and the nearest-neighbor carbon-carbon distance
$a=0.142nm$ as in a real graphene sample.\cite{ref3,ref4} The
superconductor gap $\Delta$ is set to $\Delta =t/2750 =1meV$, and
the Fermi wave-vector $k_F =1{\AA}^{-1}$. The temperature $ \cal{T}$
is set to zero since $ \cal{T}$ can be as low as 1K in a real
experiment and thus $k_B \cal{T}$ is much smaller than all other
relevant energies, such as $t$ and $\Delta$. The magnetic field is
expressed in terms of $\phi$ with $\phi \equiv (3\sqrt{3}/4) a^2
B/\phi_0$ and $(3\sqrt{3}/2) a^2 B$ is the magnetic flux in the
honeycomb lattice. In the presence of disorder, the curves are
averaged over up to 1000 random configurations.

\section{the linear conductance}

In this section, we consider the small bias limit and investigate
the linear conductance. When the bias $V$ is smaller than the gap
$\Delta$, the normal tunneling processes can not occur and
$T(\omega)=0$ for $|\omega|<\Delta$. Then only Andreev reflection
processes contribute to the current, and the linear conductance $G
=\lim\limits_{V\rightarrow 0} dI/dV = (4e^2/h) T_A(0)$ at zero
temperature. In the following, we carry out numerical studies of
graphene nanoribbons with both zigzag and armchair edges.

\subsection{The zigzag-edge case}

First, we study the clean graphene nanoribbons with the disorder
strength $W=0$. Fig.2 shows the linear conductance $G$ versus the
on-site energy $E_L$ (i.e. the energy at the Dirac point) with and
without the magnetic field. The energy $E_L$ can be controlled by
the gate voltage in an experiment. For $E_L>0$, the charge carrier
of graphene is hole-like, and it is electron-like for $E_L<0$. In
the absence of a magnetic field ($\phi=0$), the conductance $G$ is
approximately linear with $|E_L|$ due to the linear increasing of
the carrier density. For a narrow graphene nanoribbon (e.g.
$N=40$), the conductance $G$ clearly shows the step structures
because of the sub-bands from the finite width. When a sub-band
passes through the Fermi energy $E_F$ ($E_F=0$), a step appears.
For $N=40$, the width of the graphene ribbon is $(3N-1)a\approx
17nm$. The graphene ribbon with this width has been realized in a
recent experiment.\cite{ref18} On the other hand, for a wide
graphene nanoribbon (e.g. $N=70$), the step structures faint away
due to the reduction of the interval of the sub-bands.

While in the presence of a strong magnetic field, the conductance
$G$ does not show a clear pattern and depends strongly on $E_L$
and the width $N$ (see Fig.2b). Raising the disorder from zero,
the conductance $G$ in the small-value region is increased while
$G$ in the large-value region is decreased (as shown Fig.3),
meanwhile some plateaus emerge at moderate disorder strength, e.g
$W=2$.\cite{addnote21} These plateaus origin from the mixture of
the electron and hole edge states, which will be discussed in
detail in the last paragraph in this sub-section.

Fig.4 shows the linear conductance $G$ versus the on-site energy
$E_L$ with a moderate disorder strength $W$. $G$ exhibits the
plateaus with or without a magnetic field. In absence of the
magnetic field, the conductance $G$ is similar to the disorder-free
case (compare Fig.2a and Fig.4a), and the plateaus of the
conductance are equal-spaced in energy. These plateaus are from the
discrete sub-bands. In a graphene sample, due to the linear
dispersion relation, the sub-bands are equal-spaced, so are the
plateaus. For a wider graphene ribbon, the sub-bands are closer,
then the width of the plateaus are smaller, so that the plateaus are
fainted at large width (e.g. $N=70$). On the other hand, in the
presence of a magnetic field, the width of conductance plateaus are
independent to the width $N$ of the graphene ribbon, and the
plateaus are always clear regardless of $N$. Now the plateaus are
equal-spaced in the scale of $E_L^2$, and the values of the
conductance plateaus are determined by the filling factors $\nu$ of
the Landau level and the width $N$ of the graphene ribbon. The wider
$N$ is, the larger the conductance value is. But the conductance
reaches a saturated value $|\nu|e^2/h$ at large $N$ (see Fig.5b). Fig.5 shows the
conductance $G$ versus the width $N$ of the graphene ribbon. For $\phi=0$, $G$
increases approximatively in a linear way with $N$ (see
Fig.5a). But at high magnetic field $G$ has a saturated value (see
Fig.5b). For small filling factor(e.g. $E_L=0.1t$ with $\nu=2$), $G$
reaches the saturated value with small $N$ ($N=40$). For large
filling factor, $G$ reaches the saturated value only with quite
large $N$.

With the aid of the edge states, these phenomena can be well
explained. With a high magnetic field, the edge states that carry
charges are formed. In the interface of the graphene and the
superconductor, the edge states extend from one boundary to the
other along the interface, in which the Andreev reflection occurs.
So the wider the graphene ribbon is, the larger the probability is
for the Andreev reflection. In the large $N$ limit, the electron and
hole edge states are well mixed, thus, the Andreev reflection
coefficient is $0.5$, independent of any system parameters, such as
the width $N$, the on-site energy $E_L$, the magnetic field strength
$\phi$, and the disorder strength $W$. Then the conductance
$G=(2e^2/h)2T_A(0) = |\mu|e^2/h$.

\subsection{The armchair-edge case}

In this sub-section, the linear conductance $G$ in the armchair edge
case is numerically investigated. Fig.6 shows the conductance $G$
versus the on-site energy $E_L$. Without a magnetic field
($\phi=0$), $G$ increases linearly with the energy $|E_L|$ in the
absence of disorder (see Fig.6a). The disorder evidently enhances
the conductance $G$ in the small $|E_L|$ region (see Fig.6b), thus,
$G$ departs from the linear relation with $|E_L|$. In contrast with
the zigzag edge case, it has two obvious characteristics: (i) There
is a zero conductance $G$ region near $E_L=0$ for $N=3m$ or $3m+1$,
because that an energy gap emerges at the armchair edge graphene
ribbon causing the Andreev reflection to vanish. This zero
conductance $G$ region still exists in the presence of disorder
(e.g. $W=2$). (ii) The step structures from the sub-bands are not
apparent, although the width of the graphene ribbon is $\sqrt{3} Na
\approx 17nm$ for $N=70$. For the zigzag edge case with this width
the step structures are clearly seen (see Fig.2a and Fig.4a).

With a magnetic field, the Landau levels are formed and the
conductance $G$ departs completely from the linear relation with
$|E_L|$. For the clean system ($W=0$), the conductance is quite
small at the smallest filling factor $|\nu|=2$, and exhibits some
peaks at the higher filling factors $|\nu|=6$, $10$, $14$, etc (see
Fig.6c). On the other hand, in the presence of disorder ($W=2$), the
conductance $G$ shows plateaus and the plateau values are $|\nu|
e^2/h$ in the large width $N$ limit. This is because of the mixture
of the electron and hole edge states and the Andreev reflection
coefficient is $0.5$ at large $N$. The characteristics of the
plateaus at the moderate disorder strength are similar to that of
the zigzag edge graphene ribbon.

Fig.7 shows the conductance $G$ versus the width $N$ of the graphene
ribbon for a moderate disorder strength. At zero magnetic field, the
conductance $G$ increases linearly with the width $N$ as appears in
a classical system. But at a high magnetic field, although the
conductance $G$ still increases with the width, a saturation value
$|\nu| e^2/h$ appears, same as in the zigzag edge case.

\section{the finite bias case}

In this section, the case with a finite bias is investigated. With
the bias $V>\Delta$, the normal tunneling processes also occur, and
the current $I$ is:
\begin{eqnarray}
 I &=& \frac{2e}{h} \int d\omega \{ T(\omega)
 (f_{\uparrow}-f) + T_A(\omega)
 (f_{\uparrow}-f_{\downarrow}) \}.
\end{eqnarray}
Following, we numerically study the normal tunneling coefficient
$T(\omega)$ and Andreev reflection coefficient $T_A(\omega)$ for the
zigzag edge graphene ribbon. Fig.8 shows $T(\omega)$ and
$T_A(\omega)$ versus the energy $\omega$ of the incident electron
for the clean system. The normal tunneling coefficient $T(\omega)$
is zero when $|\omega|<\Delta$ because of the superconductor gap,
and $T(\omega)$ is near 1 at $|\omega|>\Delta$ since there is no
barrier at the interface of the superconductor and graphene. Next,
we focus on the Andreev reflection coefficient $T_A(\omega)$. For
zero magnetic field with $\phi=0$, $T_A(\omega)$ is almost zero for
$|\omega| >|E_L|$ (see Fig.8a), implying that the specular Andreev
reflection is very weak at $\phi=0$. But the usual Andreev
retroreflection still occurs, and $T_A(\omega)$ is quite large for
$|\omega|<|E_L|$. With a magnetic field (see Fig.8c), both kinds of
Andreev reflections occur simultaneously, and $T_A(\omega)$ is
always finite regardless whether $|\omega|<|E_L|$ or
$|\omega|>|E_L|$. $T_A(\omega)$ has a peak at $\omega=\pm\Delta$ and
quickly decays for $|\omega|>\Delta$, which is similar to a normal
metal-superconductor junction.\cite{ref24}

Finally, the effect of the disorder on the normal tunneling
coefficient $T(\omega)$ and Andreev reflection coefficient
$T_A(\omega)$ are studied. The normal tunneling coefficient
$T(\omega)$ is almost unaffected by a moderate disorder strength
$W$, $T(\omega)$ is still zero for $|\omega|<\Delta$ and near 1 for
$|\omega|>\Delta$ (see Fig.9b and 9d). However, the Andreev
reflection coefficient $T_A(\omega)$ is evidently affected by the
disorder (see Fig.9a and 9c). Both specular Andreev reflection and
the usual Andreev retroreflection occur, and $T_A(\omega)$ is close
to 0.5 in the whole range of $|\omega|<\Delta$.

\section{conclusion}

In summary, by using the non-equilibrium Green's function method,
the electron transport through the graphene
nanoribbon-superconductor junction is investigated. Both zigzag and
armchair edge graphene nanoribbons are considered. The effects of a
magnetic field and disorder on the transport property are discussed.
In the clean system and without a magnetic field, the linear
conductance increases approximatively in a linear fashion with the
on-site energy for the case with the armchair edge or the wide
zigzag edge. In the presence of a magnetic field and moderate
disorder, the linear conductance exhibits the plateau structures for
both armchair and zigzag edge nanoribbons. The plateau value
increases with the width of the graphene ribbon, but reaches a
saturation at $|\nu|e^2/h$ ($\nu$ is the filling factor) for the
wide graphene ribbon. In addition, the case with a finite bias is
studied, and the dependence of the Andreev reflection and normal
tunneling coefficients on the energy of the incident electron are
discussed.

\section*{Acknowledgments}

We gratefully acknowledge the financial support from NSF-China under
Grants Nos. 10525418 and 10734110, and US-DOE under Grants No.
DE-FG02- 04ER46124 and US-NSF.

\section*{APPENDIX}

In this appendix, we derive the surface green function ${\bf g}_S^r$
of the superconductor lead and the self-energies ${\bf
\Sigma}_R^{r,a,<}$ of coupling to the superconductor lead. The
definition of the surface green function ${\bf g}_S^r$ is:
\begin{eqnarray}
 {\bf g}_S^r(y,y',t) = -i\theta(t) \left(\begin{array}{ll}
 \langle \{b_{\uparrow}(y,t), b^{\dagger}_{\uparrow}(y',0) \}\rangle
 &
 \langle \{b_{\uparrow}(y,t), b_{\downarrow}(y',0) \}\rangle \\
 \langle \{b^{\dagger}_{\downarrow}(y,t), b^{\dagger}_{\uparrow}(y',0) \}\rangle
 &
 \langle \{b^{\dagger}_{\downarrow}(y,t), b_{\downarrow}(y',0) \}\rangle
 \end{array} \right) \nonumber
\end{eqnarray}
and ${\bf g}_S^r(y,y',\omega)$ is the Fourier transformation of
${\bf g}_S^r(y,y',t)$, where $y$ and $y'$ are the real-space
positions on the surface of half-infinite superconductor lead.
Applying the equation of motion, ${\bf g}_S^r(y,y',\omega)$ can be
written as:
\begin{eqnarray}
 {\bf g}_S^r(y,y',\omega) = \sum\limits_{\bf k}
 \frac{1}{\omega_+^2-\epsilon_{\bf k}^2 -\Delta^2}
 \left( \begin{array}{ll}
 \omega_+ +\epsilon_{\bf k} & \Delta \\
 \Delta & \omega_+ -\epsilon_{\bf k}
  \end{array}    \right)
 e^{i k_y (y-y')}  \nonumber,
\end{eqnarray}
where $\omega_+ =\omega +i0^+$. Next, we calculate the sum,
$\sum\limits_{\bf k} F({\bf k}) e^{ik_y(y-y')}$ with $ F({\bf k}) =
(\omega_+ \pm \epsilon_{\bf k})/(\omega_+^2 -\epsilon_k^2
-\Delta^2)$ or $ F({\bf k}) = \Delta/(\omega_+^2 -\epsilon_k^2
-\Delta^2)$,
\begin{eqnarray}
\sum\limits_{\bf k} F({\bf k}) e^{ik_y(y-y')} & =& \int_{-\pi}^{\pi}
d\theta \int
dk k \rho_k e^{ik(y-y')\sin \theta} F(k) \nonumber\\
& =& \int d\epsilon_{k} J_0(k(y-y')) \rho(\epsilon_k) F(k).\nonumber
\end{eqnarray}
where $J_0$ is the Bessel function of the first kind, $\rho_k$ is
the density of state in the $k$ space and $\rho(\epsilon_k) = 2\pi k
\rho_k (dk/d\epsilon_k)$ is the density of state in the energy
space. In the above steps, we have assumed the s-wave superconductor
so that $\epsilon_{\bf k}$ only depends on $k=|{\bf k}|$. In the
following, we assume that the density of state
$\rho(\epsilon_k)=\rho$ is independent of the energy $\epsilon_k$
and $J_0(k(y-y'))$ only depends on the Fermi wave-vector $k_F$.\cite{ref22}
These assumptions are reasonable because that the main contribution
to the transport behaviors is these electrons with their energies
near the Fermi energy. Then ${\bf g}_S^r(y,y',\omega)$ reduces to:
\begin{eqnarray}
 {\bf g}_S^r(y,y',\omega) = J_0(k_F(y-y')) \rho \int d\epsilon_k
 \frac{1}{\omega_+^2-\epsilon_{k}^2 -\Delta^2}
 \left( \begin{array}{ll}
 \omega_+ +\epsilon_{k} & \Delta \\
 \Delta & \omega_+ -\epsilon_{k}
  \end{array}    \right) \nonumber.
\end{eqnarray}
By using the theorem of residue, the integration $\int d\epsilon_k$
in the above equation can be obtained analytically,\cite{ref22} and
the surface Green's function ${\bf g}_S^r(y,y',\omega)$ changes
into:
\begin{eqnarray}
 {\bf g}_S^r(y,y',\omega) = -i\pi \rho J_0(k_F(y-y')) \beta(\omega)
 \left( \begin{array}{ll}
 1  & \Delta/\omega \\
 \Delta/\omega & 1
  \end{array}    \right) \nonumber,
\end{eqnarray}
where $\beta(\omega) =|\omega|/\sqrt{\omega^2-\Delta^2}$ while
$|\omega|>\Delta$ and $\beta(\omega) =
\omega/(i\sqrt{\Delta^2-\omega^2})$ while $|\omega| <\Delta$. After
solving the surface Green's function ${\bf g}_S^r(y,y',\omega)$, the
self-energies ${\bf \Sigma}_R^{r,a,<}$ are obtained
straightforwardly.
\begin{eqnarray}
 {\bf \Sigma}_{R,ij}^r(\omega) & = & t_c {\bf g}_S^r(y_i,y_j,\omega)
 t_c^* \nonumber\\
 & =& -i\pi \rho |t_c|^2 J_0(k_F(y_i-y_j)) \beta(\omega)
 \left( \begin{array}{ll}
 1  & \Delta/\omega \\
 \Delta/\omega & 1
  \end{array}    \right) \nonumber\\
 & \equiv & -(i/2) {\bf \Gamma}_{R,ij}(\omega), \nonumber
\end{eqnarray}
${\bf \Sigma}_{R}^a(\omega) = (i/2){\bf \Gamma}_{R}(\omega)$ and
${\bf \Sigma}_{R}^<(\omega) = if(\omega){\bf \Gamma}_{R}(\omega)$.

\newpage

\begin{figure}
\caption{ (a) and (b) are the schematic diagrams for the zigzag and
armchair edge graphene nanoribbon-superconductor junctions,
respectively. } \label{fig:1}

\caption{ (Color online) The linear conductance $G$ vs. the energy
$E_L$ for different width $N$ at the clean system with $W=0$. The
panels (a) and (b) are for the magnetic field strength $\phi=0$ and
$\phi=0.007$, respectively. } \label{fig:2}

\caption{ (Color online) The conductance $G$ vs. $E_L$ for the
different disorder strengths $W$, with the parameters $L=16$,
$N=60$, and $\phi=0.007$.
 } \label{fig:3}

\caption{ (Color online) The conductance $G$ vs. the energy $E_L$
for different width $N$ at the moderate disorder strength $W=2t$ and
$L=16$. The panels (a) and (b) are for the magnetic field strength
$\phi=0$ and $\phi=0.007$, respectively. } \label{fig:4}

\caption{ (Color online) The conductance $G$ vs. the width $N$ of
the graphene nanoribbon with $E_L=0.1t$ (solid curve), $0.2t$
(dashed curve), and $0.25t$ (dotted curve). The panels (a) and (b)
are for the magnetic field strengths $\phi=0$ and $\phi=0.007$,
respectively. The other parameters are $W=2$ and $L=16$. }
\label{fig:5}

\caption{ (Color online) The conductance $G$ vs. the energy $E_L$
for different width $N$. The parameters are $L=12$, the disorder
strength $W=0$ [in (a) and (c)] and $W=2t$ [in (b) and (d)], and the
magnetic field strength $\phi=0$ [in (a) and (b)] and $\phi=0.007$
[in (c) and (d)]. } \label{fig:6}

\caption{ (Color online) The conductance $G$ vs. the width $N$ of
the graphene nanoribbon for the different energy $E_L$. The panels
(a) and (b) are for the magnetic field strengths $\phi=0$ and
$\phi=0.007$, respectively. The other parameters are $W=2t$ and
$L=12$.} \label{fig:7}

\caption{ (Color online) The normal tunneling coefficient
$T(\omega)$ [in (b) and (d)] and Andreev reflection coefficient
$T_A(\omega)$ [in (a) and (c)] vs. $\omega$ for different on-site
energy $E_L$. The other parameters are $N=50$, $W=0$, and the
magnetic field strength $\phi=0$ [in (a) and (b)] and $\phi=0.007$
[in (c) and (d)]. } \label{fig:8}

\caption{ (Color online) The normal tunneling coefficient
$T(\omega)$ [in (b) and (d)] and Andreev reflection coefficient
$T_A(\omega)$ [in (a) and (c)] vs. $\omega$ for different on-site
energy $E_L$. The other parameters are $N=50$, $L=16$, $W=2t$, and
the magnetic field strength $\phi=0$ [in (a) and (b)] and
$\phi=0.007$ [in (c) and (d)]. } \label{fig:9}
\end{figure}


\begin{references}
\bibitem[*]{} Electronic address: sunqf@aphy.iphy.ac.cn

\bibitem{ref1}
K.S. Novoselov, A.K. Geim, S.V. Morozov, D. Jiang, Y. Zhang, S.V.
Dubonos, I.V. Grigorieva, and A.A. Firsov, Science {\bf 306}, 666
(2004); K.S. Novoselov, A.K. Geim, S.V. Morozov, D. Jiang, M.I.
Katsnelson, I.V. Grigorieva, S.V. Dubonos, and  A.A. Firsov, Nature
(London) {\bf 438}, 197 (2005).

\bibitem{ref2}
Y. Zhang, Y.-W. Tan, H.L. Stormer, and P. Kim, Nature (London) {\bf
438}, 201 (2005).

\bibitem{ref3}
C.W.J. Beenakker, Rev. Mod. Phys. {\bf 80}, 1337 (2008)

\bibitem{ref4}
A.H. Castro Neto, F.Guinea ,N.M.R. Peres ,K.S. Novselov, and A.K.
Geim, Rev. Mod. Phys. {\bf 81}, 109 (2009).

\bibitem{ref5}
C.W.J. Beenakker, Phys. Rev. Lett. {\bf 97}, 067007 (2006).

\bibitem{ref6}
A.R. Akhmerov and C.W.J. Beenakker, Phys. Rev. B {\bf 75}, 045426
(2007).

\bibitem{ref7}
J. Linder and A. Sudb$\phi$, Phys. Rev. Lett. {\bf 99}, 147001
(2007); Phys. Rev. B {\bf 77}, 064507 (2008).

\bibitem{ref8}
M. Titov and C.W.J. Beenakker, Phys. Rev. B {\bf 74}, 041401 (2006);
M. Titov, A. Ossipov, and C.W.J. Beenakker, Phys. Rev. B {\bf 75},
045417 (2007).

\bibitem{ref9}
A.G. Moghaddam and M. Zareyan, Phys. Rev. B {\bf 74}, 241403(R)
(2006).

\bibitem{ref9n}
Q. Liang, Y. Yu, Q. Wang, and J. Dong, Phys. Rev. Lett. {\bf 101},
187002 (2008)

\bibitem{ref10}
Q. Zhang, D. Fu, B. Wang, R. Zhang, and D.Y. Xing, Phys. Rev. Lett.
{\bf 101}, 047005 (2008); J. Linder, T. Yokoyamn, D.
Huertas-Hernando, and A. Sudb$\phi$, Phys. Rev. Lett. {\bf 100},
187004 (2008); Y. Asano, T. Yoshida, Y. Tanaka, and A.A. Golubov,
Phys. Rev. B {\bf 78}, 014514 (2008).

\bibitem{ref11}
S. Bhattacharjee and K. Sengupta, Phys. Rev. Lett. {\bf 97}, 217001
(2006); S. Bhattacharjee, M. Maiti, and K. Sengupta, Phys. Rev. B
{\bf 76}, 184514 (2007).

\bibitem{ref12}
M. Maiti and K. Sengupta, Phys. Rev. B {\bf 76}, 054513 (2007).

\bibitem{ref13}
P. Burset, A.L. Yeyati, and A. Martin-Rodero, Phys. Rev. B {\bf 77},
205425 (2008).

\bibitem{ref14}
J.C. Cuevas and A.L. Yeyati, Phys. Rev. B {\bf 74}, 180501(R)
(2006).

\bibitem{aref1}
Z.Y. Zhang, J. Phys.: Condens. Matter {\bf 20}, 445220 (2008); C.
Bai, Y. Yang and X. Zhang, J. Phys.: Condens. Matter {\bf 20},
335202 (2008).

\bibitem{aref2}
J. Gonzalez and E.X. Perfetto, J. Phys.: Condens. Matter {\bf 20},
145218 (2008).

\bibitem{ref15}
A.F. Andreev, Sov. Phys. JETP {\bf 19}, 1228 (1964).

\bibitem{ref16}
H.B. Heersche, P. Jarillo-Herrero, J.B. Oostinga, L.M.K.
Vandersypen, and A.F. Morpurgo, Nature {\bf 446}, 56 (2007).

\bibitem{ref17}
F. Miao, S. Wijeratne, Y. Zhang, U.C. Coskun, W. Bao, and C.N. Lau,
Science {\bf 317}, 1530 (2007).

\bibitem{ref18}
X. Li, X. Wang, L. Zhang, S. Lee, and H. Dai, Science {\bf 319},
1229 (2008).

\bibitem{ref19}
D.N. Sheng, L. Sheng, and Z.Y. Weng, Phys. Rev. B {\bf 73}, 233406
(2006); Z. Qiao and J. Wang, Nanotechnology {\bf 18}, 435402 (2007).

\bibitem{ref20}
W. Long, Q.-F. Sun, and J. Wang, Phys. Rev. Lett. {\bf 101}, 166806
(2008); J. Li and S.-Q. Shen, Phys. Rev. B {\bf 78}, 205308 (2008).

\bibitem{ref21}
Y. Meir and N. S. Wingreen, Phys. Rev. Lett. {\bf 68}, 2512 (1992);
A.-P. Jauho, N.S. Wingreen, and Y. Meir, Phys. Rev. B {\bf 50}, 5528
(1994).

\bibitem{ref22}
Q.-F. Sun, J. Wang, and T.-h. Lin, Phys. Rev. B {\bf 59}, 3831
(1999); Phys. Rev. B {\bf 59}, 13126 (1999).

\bibitem{ref23}
D.H. Lee and J.D. Joannopoulos, Phys. Rev. B {\bf 23}, 4997 (1981);
M.P. Lopez Sancho, J.M. Lopez Sancho, and J. Rubio, J. Phys. F: Met.
Phys. {\bf 14}, 1205 (1984); {\bf 15}, 851 (1985).

\bibitem{addnote21}
In a recent paper (I.L. Aleiner and K.B. Efetov, Phys. Rev. Lett.
{\bf 97}, 236801 (2006)), it shows that the Anderson localization
occurs in the graphene systems and the conductance is very small
at the disorder strength $W\sim t$. But in this work, we only
consider a graphene nanoribbon with small sizes, so the Anderson
localization may not appear and the conductance can be large at
the disorder strength $W\sim t$.

\bibitem{ref24}
G.E. Blonder, M. Tinkham, and T.M. Klapwijk, Phys. Rev. B {\bf 25},
4515 (1982).

\end{references}
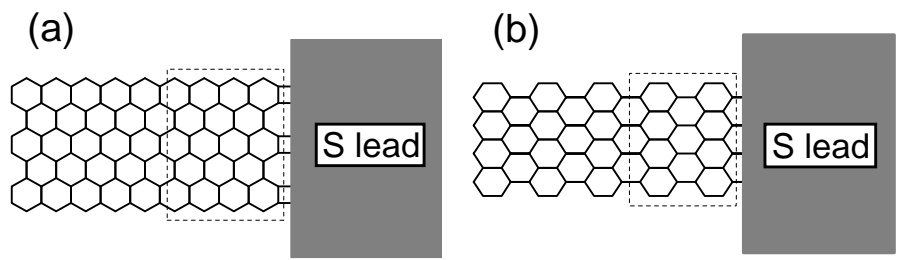
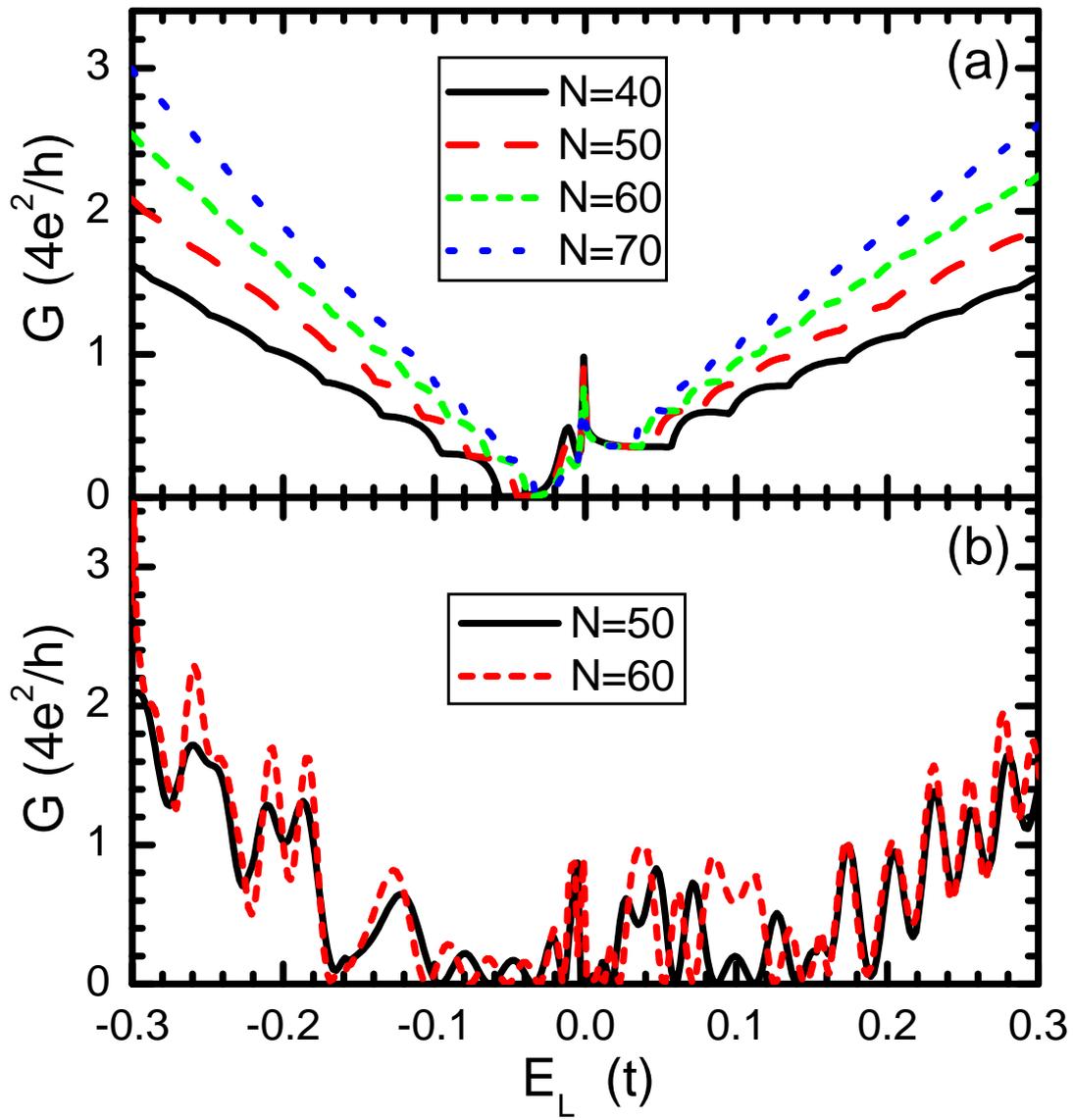
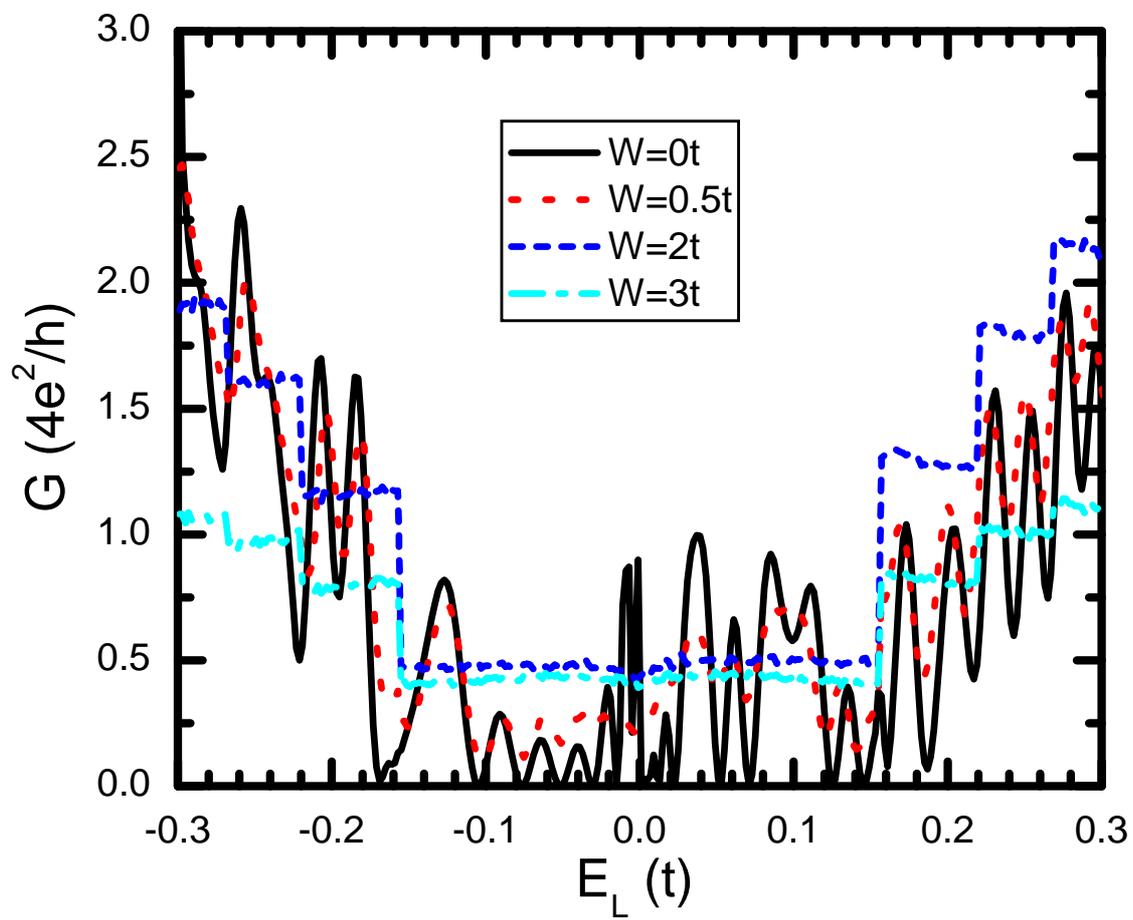
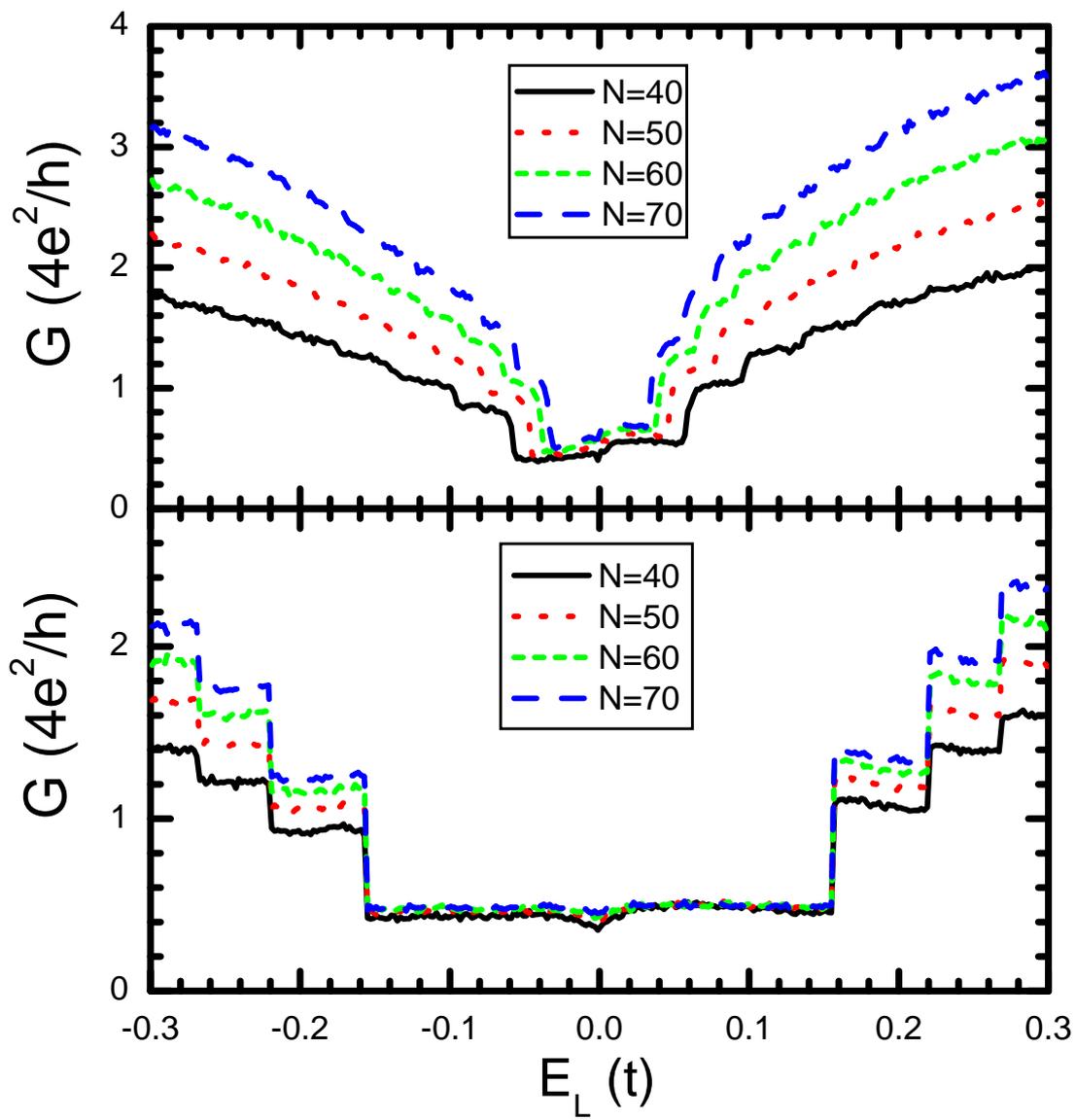
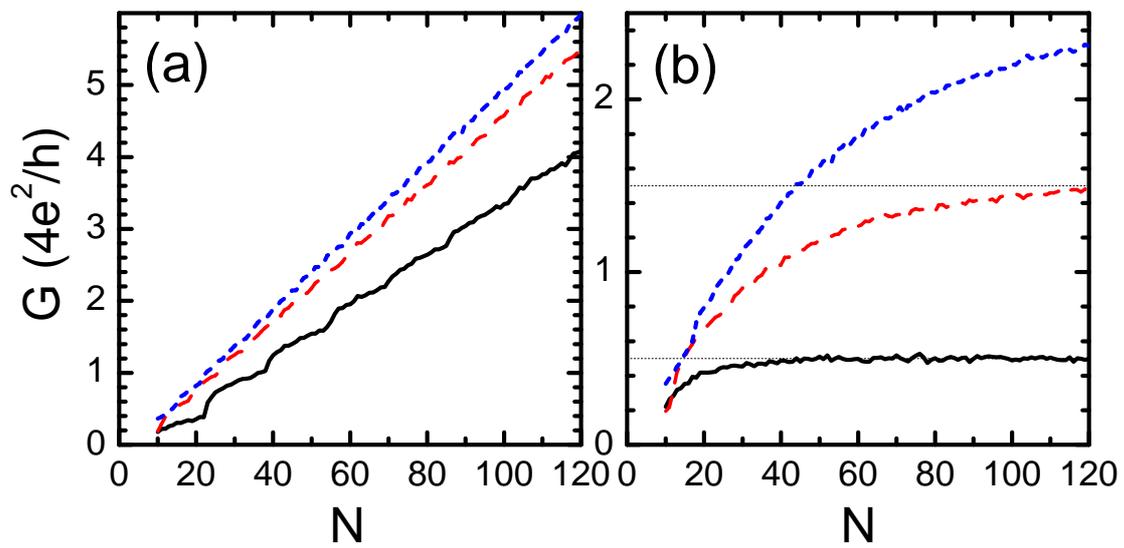
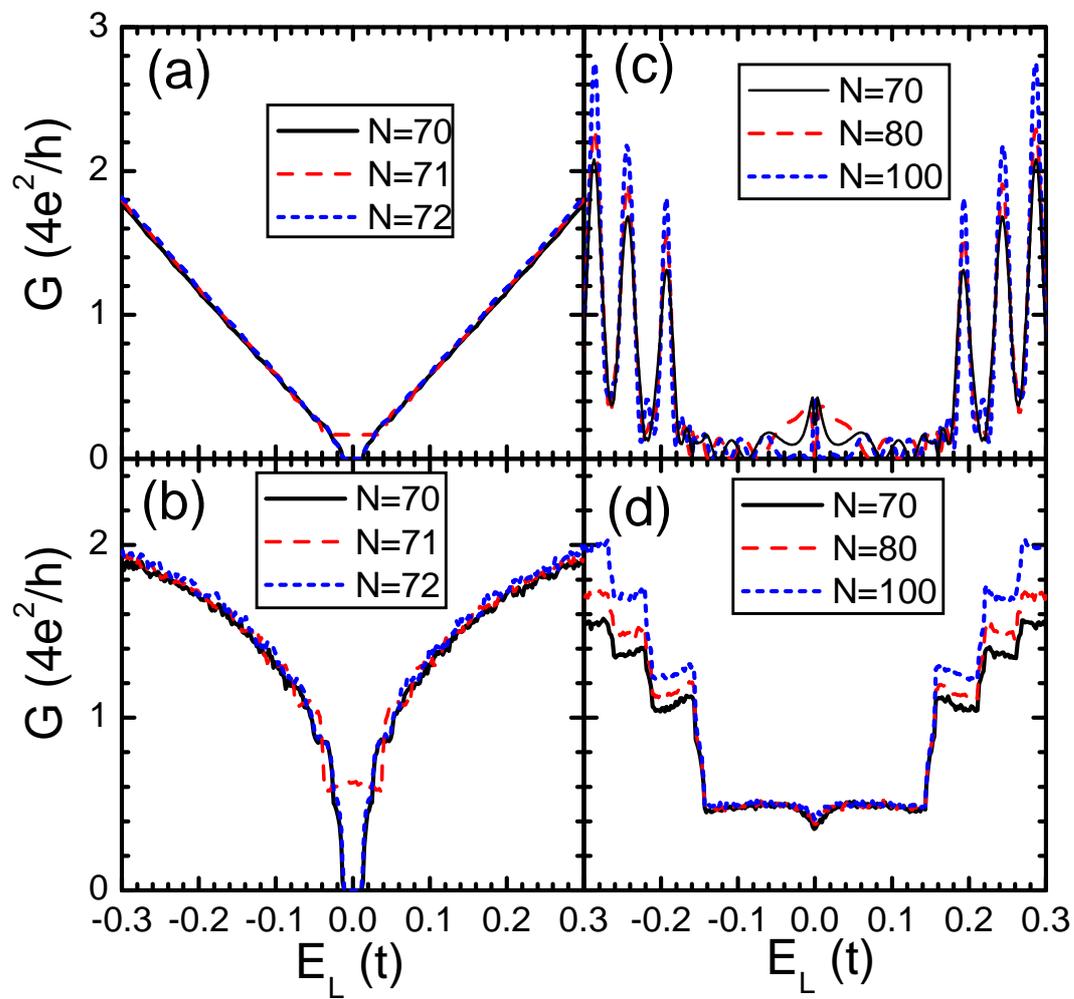
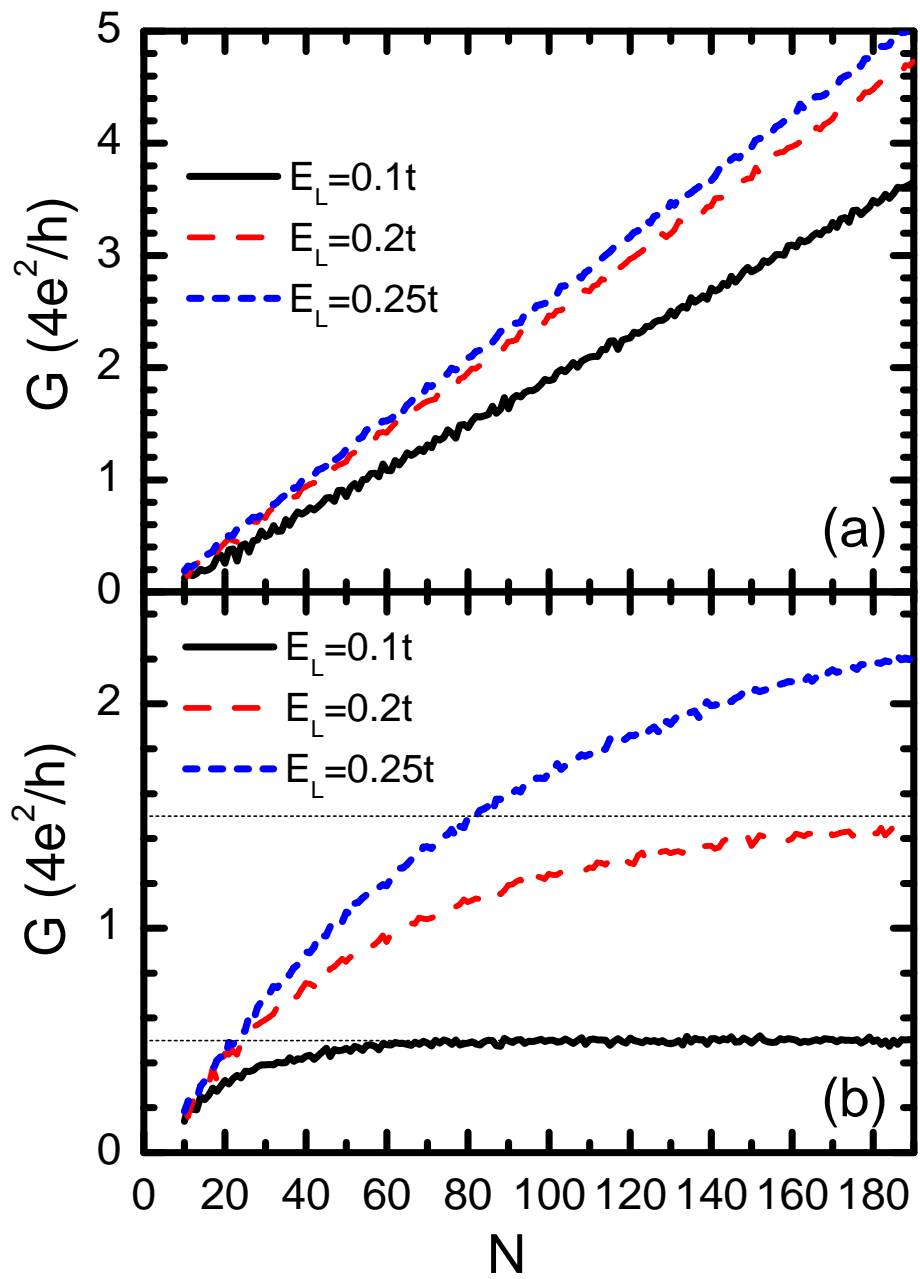
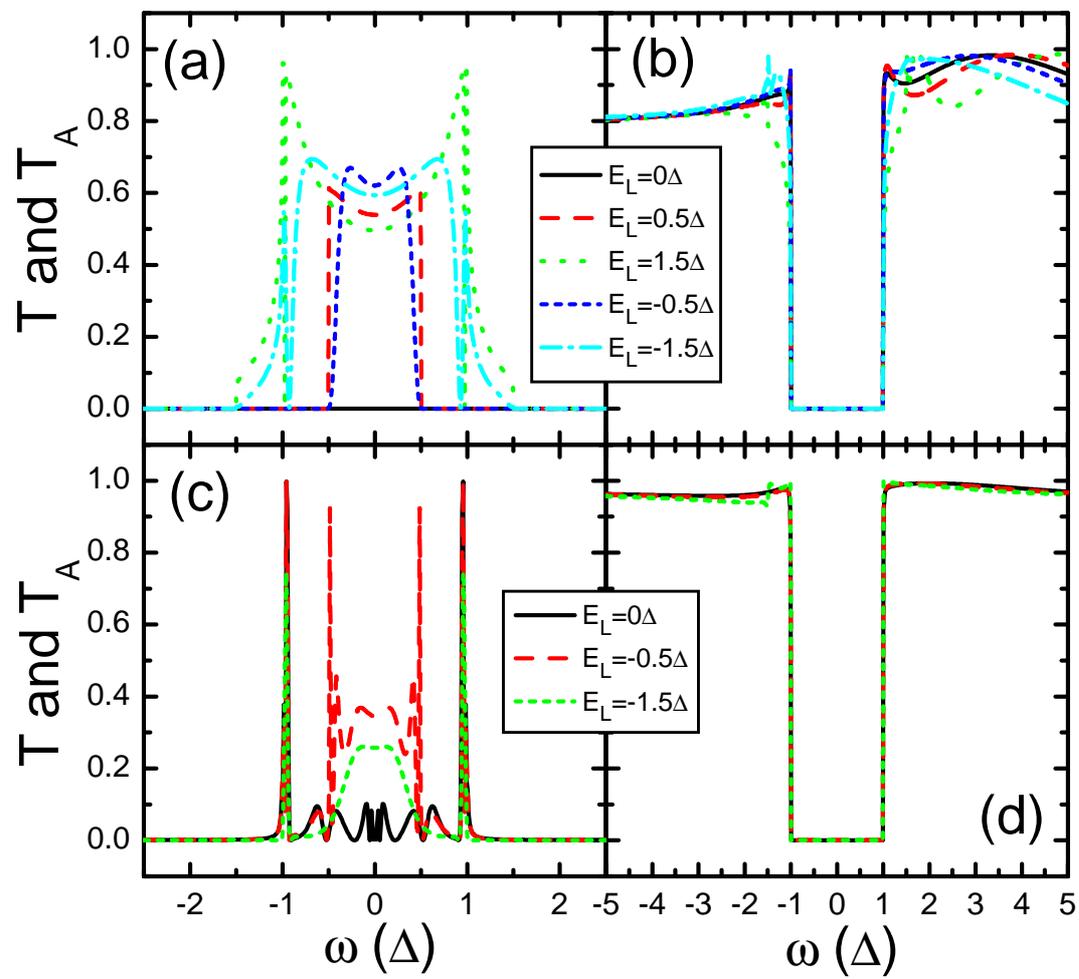
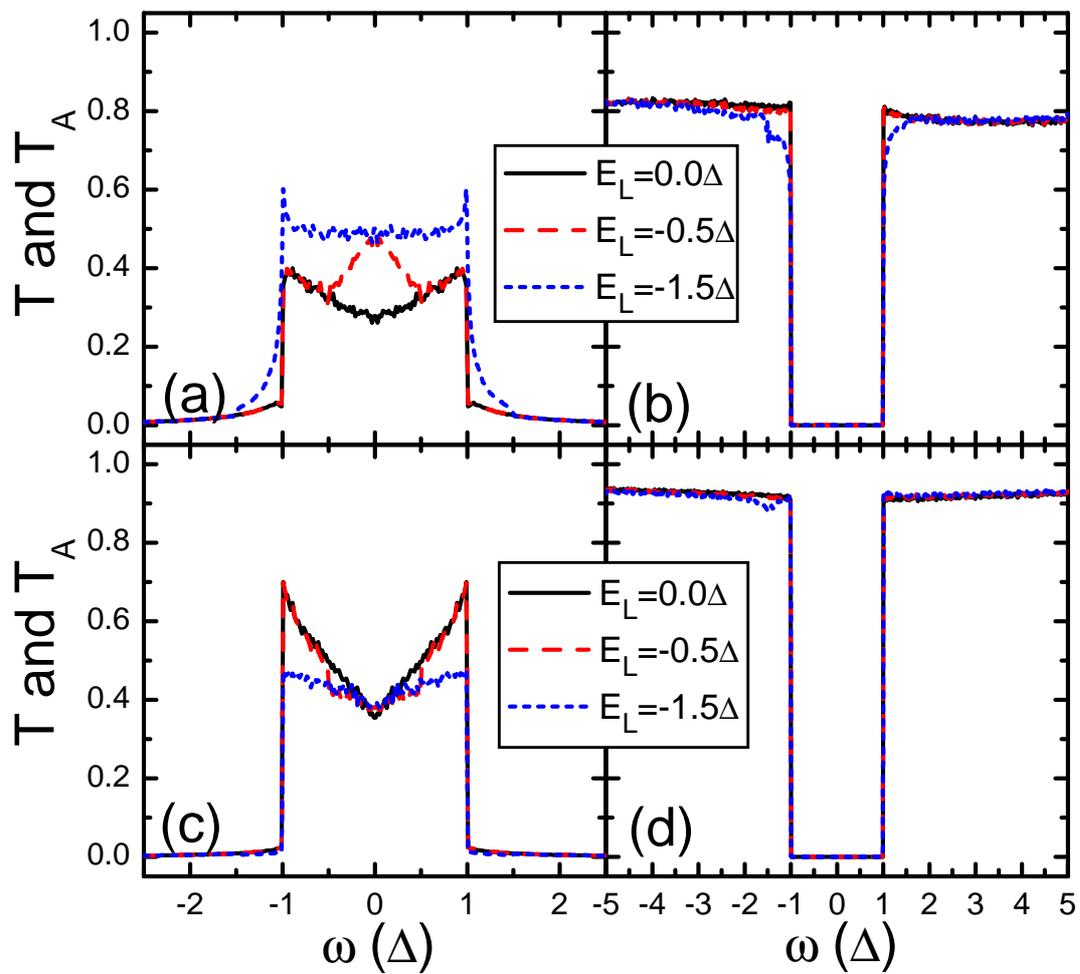
\end{document}